%% file: main.tex
\documentclass[sigconf,screen,authorversion]{acmart}
\usepackage{listings}

\AtBeginDocument{%
  \providecommand\BibTeX{{%
    \normalfont B\kern-0.5em{\scshape i\kern-0.25em b}\kern-0.8em\TeX}}}

\setcopyright{acmlicensed}
\acmPrice{15.00}
\acmDOI{10.1145/3573382.3623706}
\acmYear{2023}
\copyrightyear{2023}
\acmSubmissionID{playcomp23int-p1013-p}
\acmISBN{979-8-4007-0029-3/23/10}
\acmConference[CHI PLAY Companion '23]{Companion Proceedings of the Annual Symposium on Computer-Human Interaction in Play}{October 10--13, 2023}{Stratford, ON, Canada}
\acmBooktitle{Companion Proceedings of the Annual Symposium on Computer-Human Interaction in Play (CHI PLAY Companion '23), October 10--13, 2023, Stratford, ON, Canada}
\received{2023-06-22}
\received[accepted]{2023-08-03}

\begin{document}
\title{Lottery and Sprint: Generate a Board Game with Design Sprint Method on AutoGPT}

\author{Maya Grace Torii}
\authornote{Both authors contributed equally to this research.}
\orcid{0000-0003-4025-9212}
\email{toriparu@digitalnature.slis.tsukuba.ac.jp}
\affiliation{%
  \institution{University of Tsukuba}
  \city{Tsukuba}
  \country{Japan}
}

\author{Takahito Murakami}
\authornotemark[1]
\orcid{0000-0003-2077-9747}
\email{takahito@digitalnature.slis.tsukuba.ac.jp}
\affiliation{%
  \institution{University of Tsukuba}
  \city{Tsukuba}
  \country{Japan}
}

\author{Yoichi Ochiai}
\orcid{0000-0002-4690-5724}
\email{wizard@slis.tsukuba.ac.jp}
\affiliation{%
  \institution{University of Tsukuba}
  \city{Tsukuba}
  \country{Japan}
}



\renewcommand{\shortauthors}{Torii and Murakami, et al.}

\begin{abstract}
  In this paper, we introduce "Lottery and Sprint", a board game creation methodology that cooperates human design intuition with the structured Design Sprint framework executed by the AutoGPT system. By aligning AI-driven processes with human creativity, we aim to facilitate a collaborative game design experience. A user study is conducted to investigate the playability and enjoyment of the generated games, revealing both successes and challenges in employing systems like AutoGPT for board game design. Insights and future research directions are proposed to overcome identified limitations and enhance computational-driven game creation.
\end{abstract}

\ccsdesc[300]{Human-centered computing~Interactive systems and tools}
\ccsdesc[500]{Applied computing~Arts and humanities}

\keywords{board game design, auto game creation, design sprint, generative AI}



\maketitle
\section{Introduction}
\input{camera-ready_sections/1_introduction}
\section{Related Works}
\input{camera-ready_sections/2_relatedwork}
\section{System Overview}
\input{camera-ready_sections/3_system}
\section{User Study}
\input{camera-ready_sections/4_userstudy}
\section{Result and Discussion}
\input{camera-ready_sections/5_result-discuss}
\section{Limitation and Future Work}
\input{camera-ready_sections/6_limitation-future}
\section{Conclusion}
\input{camera-ready_sections/7_conclusion}

\begin{acks}
The manuscript was drafted using OpenAI ChatGPT and GPT-4. The AI generated text was read, revised and proofed carefuly by the authors. This work was supported by Grant-in-Aid for JSPS Fellows Grant Number JP23KJ0265.
\end{acks}

\bibliographystyle{ACM-Reference-Format}
\bibliography{camera-ready_ref}


\end{document}

%% file: camera-ready_sections/1_introduction.tex
Creating novel, enjoyable and effective board games typically requires a detailed understanding of game mechanics, player engagement, and strategic balance~\cite{eck2017leveling}. Inexperienced individuals often face challenges in designing board games that cater to various play styles, objectives, and constraints~\cite{book}. To address this, we present the "Lottery and Sprint" method. This approach allows human designers to work with the AutoGPT system—an AI model based on Generative Pre-trained Transformers (GPT)—wherein the Design Sprint framework is embedded into the AutoGPT's prompt mechanism. Through this method, the collaboration between human and AI leads to the creation of innovative and balanced board games.

This research explores the application of the AutoGPT system in board game design, detailing our "Lottery and Sprint" approach. The process begins with AutoGPT generating a variety of game designs in a metaphorical "lottery", allowing users to select a favored design. Once a game is chosen, it enters a customized Design Sprint~\cite{knapp2016sprint} process to refine and improve the game concept. The central research question of this study is whether the AutoGPT system, operating within the proposed framework, can effectively create board games that are enjoyable, playable, and captivating, even for individuals with limited experience in game design.

Our investigation encompasses a review of related works. We then present the system design using AutoGPT and customized Design Sprint. This is followed by a user study evaluating the generated board games in terms of playability and entertainment by Creativity Support Index (CIS) and qualitative analysis. 

%% file: camera-ready_sections/2_relatedwork.tex
Game creation tools have advanced significantly in recent years, with techniques like Monte Carlo tree search enabling automated game content generation~\cite{chaslot2008monte}. However, these systems often focus on automation and may not effectively support human designers' creative ideation. For instance, GDL~\cite{thielscher2017gdl} employs complex formalisms that could hinder non-experts' intuitive expression of ideas~\cite{ludii}, while Monte Carlo tree search in Regular Boardgames~\cite{kowalski2019regular} primarily targets AI techniques rather than assisting designers. As a result, there is room for improved human-computer collaboration in the game design process.

In this context, AI-driven game design has introduced new possibilities for collaborative game creation using AI~\cite{khalifa2019general}. 

The rapid growth and advancements in GPT and Large Language Models (LLMs) have expanded creative processes across various domains~\cite{yang2023harnessing}. For example, Buncho employs GPT-2 to enhance writers' enjoyment in crafting novels~\cite{buncho}, and similar capabilities have been applied to adjust video game level difficulty~\cite{10.1145/3582437.3587211}.

Considering these developments, AutoGPT, an open-source project for automatic task decomposition and problem-solving, emerges as a promising candidate for creative endeavors~\cite{Auto-GPT}. Yang et al. conducted a comprehensive benchmark study on AutoGPT agents in real-world scenario simulations~\cite{yang2023autogpt}. Building upon this foundation, our study aims to explore AutoGPT's potential in board game design through the "Lottery and Sprint" method. This approach seeks to facilitate collaboration between human designers and AI, with the goal of creating engaging and playable board games for individuals with limited design experience while maintaining a rigorous analytical perspective.

%% file: camera-ready_sections/3_system.tex
\begin{figure*}[bht]
  \centering
  \includegraphics[width=\linewidth]{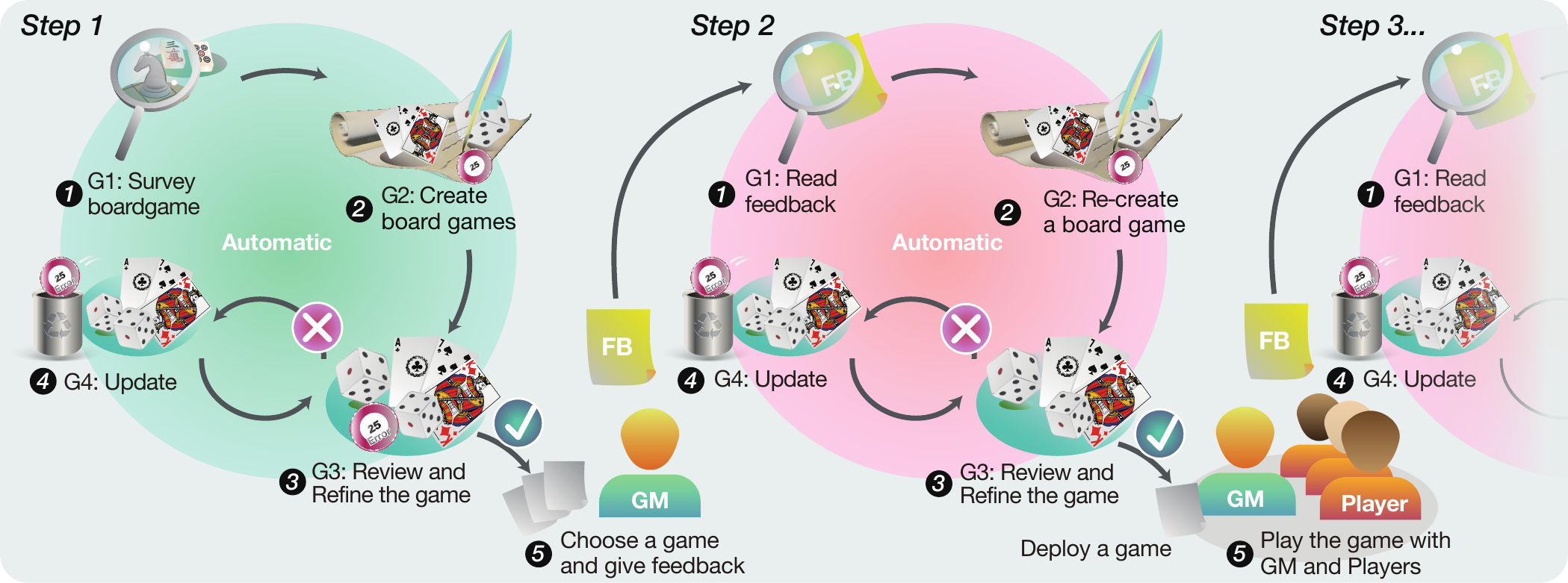}
  \caption{Overview of the "Lottery and Sprint" process, which is inspired by the Design Sprint framework. Step 1, the Lottery phase (Prompt ~\ref{lst:1}), consists of (1) G1: AutoGPT surveying online board games, (2) G2: AutoGPT creating board games based on user constraints, (3) G3: Reviewing and refining the generated games, and (4) G4: Updating the board game according to the refining process. This process follows a  cyclical pattern, inspired by the flexibility and iterative aspects of Design Sprint, and continues until the generated rules are bug-free. Then, in G5, the single user (GM) selects a game from the generated options then gives feedback to AutoGPT (Prompt ~\ref{lst:2}). In Step 2, the process follows a similar pattern, with (1) G1: GM reading user feedback in detail, (2) G2: AutoGPT regenerating the selected game based on feedback provided by the GM, (3) G3: Reviewing and refining the game again, and (4) G4: Updating the board game according to the refining process. This cycle repeats until the game is ready for test play (G5) with multiple user groups (GM and players). In Step 3 and onward, the process repeats the pattern inspired by Design Sprint as in Step 2, with feedback from test play being incorporated into the game by the GM, leading to further refinement and updates with the help of AutoGPT.}
  \Description{}
  \label{so}
\end{figure*}

\begin{figure*}[htb]
  \centering
  \includegraphics[width=\linewidth]{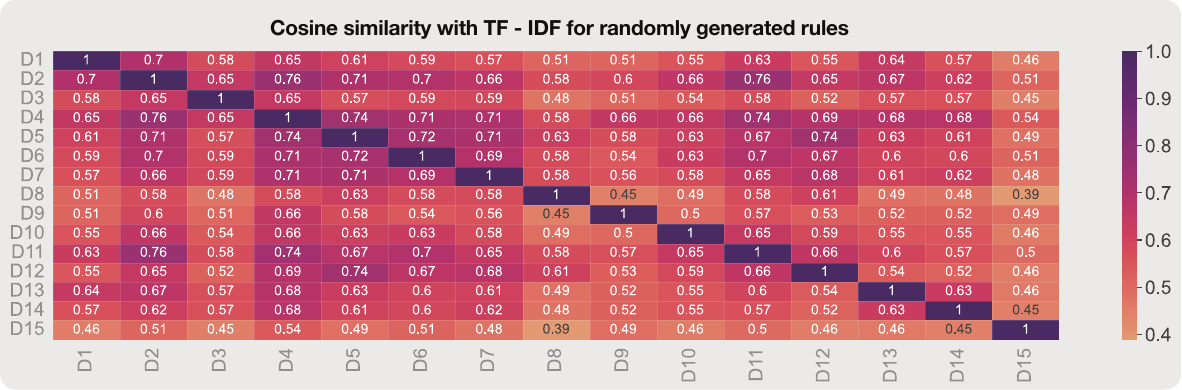}
  \caption{Using the creation prompt, 15 board games were randomly generated in the same condition. The generated text was then evaluated through a TF-IDF test, followed by Cosine Similarity, and the results are presented.}
  \Description{Using the creation prompt, 15 board games were randomly generated in the same condition. The generated text was then evaluated through a TF-IDF test, followed by Cosine Similarity, and the results are presented.}
  \label{qualitive}
\end{figure*}

\subsection{Design Sprint and Its Integration}
Design Sprint is a time-bound, structured process that aims to solve complex problems through design, prototyping, and testing ideas with users. Developed by Google Ventures, it is a five-day process where each day is dedicated to a specific phase: Understand, Diverge, Converge, Prototype, and Test~\cite{knapp2016sprint, inbook}. In the context of the AutoGPT board game creation system, the principles of Design Sprint are integrated into the architecture, enhancing the efficiency and effectiveness of the game design process. This integration can be mapped onto the four key stages of the system's architecture as follows:

1. Research existing board games (Understand and Diverge): In this stage, the system surveys existing board games, gathering insights and generating ideas, similar to the Understand and Diverge phases of Design Sprint. The system learns about diverse rule sets, game balance, and strategic depth to inform its game design process.
2. Create a board game with constraints (Converge and Prototype): This stage corresponds to the Converge and Prototype phases of Design Sprint. The system uses the insights gathered from the research stage to create a new board game based on given materials and constraints, akin to building a prototype for testing.
3. Reflect on game design (Test): The Test phase of Design Sprint is mirrored in this stage, where the system evaluates the created game based on usability, rule clarity, design consistency, strategic balance, and enjoyment. This serves as an internal evaluation mechanism for the system’s output.
4. Update and iterate (Iterate): The final stage of the AutoGPT system emphasizes iteration, a core principle of Design Sprint. Informed by the reflections from the previous stage, the system refines and updates the game to ensure that all materials, instructions, and game elements are clear and consistent.

\subsection{Prompts}
In the AutoGPT board game creation system(Fig.~\ref{so}) \footnote[1]{\url{https://github.com/DigitalNatureGroup/Lottery-and-Sprint}\label{github}}, the four key stages of the architecture are supported by two types of prompts, creation prompts and feedback prompts. These prompts guide the AutoGPT system during the design process and are divided according to their role within the system architecture.
Creation Prompts play a role in the first step of the system architecture: they survey, create an initial draft of board game rules, reflect on the created game, and update the game based on that reflection.
\vspace{2mm}
\begin{lstlisting}[caption= Creation prompts, label={lst:1}]
(G1) In order to make an interesting board game, survey board games focusing on basic various rules of board game, how other games are keeping game balance, and how other games exist.
(G2) Create a board game with the material and constraint written in text file "<board game material and constraint>.txt". Output the game explanation including all information to play the game with Title, Materials, setup, gameplay rules, game ending conditions, board layout and design, the unique point, the enjoyable point and strategy to win the game as text file so you can memorize or refer any time.
(G3) Step-by-step reflect the game you have created and consider if there are any failure in following the materials and constraints, difficult instructions for human to keep track while playing the game(usability in games), unclear rule instruction for situation potentially to occur(rule design), failure in rule, inconsistent in between the winning strategy and intention of the game(board game design and game balance, game theory) and export your result in text file. If there are not any problems, write so. use_gpt4 for better reflection.
(G4) Reading the reflection, updating the game. Check if the game is perfect and ready to play by taking the step in (G3).
\end{lstlisting}
\vspace{2mm}
Feedback Prompts are involved in the later stages of the system architecture. In these stages, the system processes user feedback and updates the game accordingly: it reads the feedback, recreates the board game, reflects on the game, and then updates it.
\vspace{2mm}
\begin{lstlisting}[caption= Feedback prompts, label={lst:2}]
(G1) Read the board game rule which was created and written in "<Target board game>.txt", material and constraints defined in "<board game material and constraint>.txt".
(G2) You have feedback from players of the game in this file "<feedback>.txt". Read the file and recreate the game. output and save the reflection to the text file so you can refer any time.
(G3) Step-by-step reflect the game you have created by considering the materials used (only defined materials should be used), rules or instructions to be very clear for first time players to play, clear winning conditions. output and save the reflection to the text file so you can refer any time.
(G4) Reading the reflection, updating the game by cooperating with the reflections and output and saving the updated game to a text file so you can refer any time. Make sure to include all the rules in the text file.
\end{lstlisting}
\vspace{2mm}

\subsection{System Output and Evaluation}
As defined in the prompts (Prompt~\ref{lst:1}), the system generates board game rules including elements such as title, materials, setup, gameplay rules, game ending conditions, etc\footnotemark[1].

Our analysis shown in Fig.~\ref{qualitive} evaluate the diversity of game rules generated by the lottery phase using the AutoGPT system. We generated 15 different game rules and analysed their similarity using NLTK for morphological analysis~\cite{loper2002nltk} and the TF-IDF method for text vector representation~\cite{singhal2001modern}.

By using cosine similarity, which ranges from 0 to 1, to compare the TF-IDF representations of the game rules, we were able to evaluate the diversity of the generated content over randomness. Our results show a mean cosine similarity of 0.59, a standard deviation of 0.08, with minimum and maximum cosine similarity values of 0.39 and 0.76 respectively. These values suggest that there is at least a reasonable level of diversity in the proposed designs, given the limitations of the data and constraints provided to users, such as the limited number of players and items available.

%% file: camera-ready_sections/4_userstudy.tex
\begin{figure*}[ht]
  \centering
  \includegraphics[width=\linewidth]{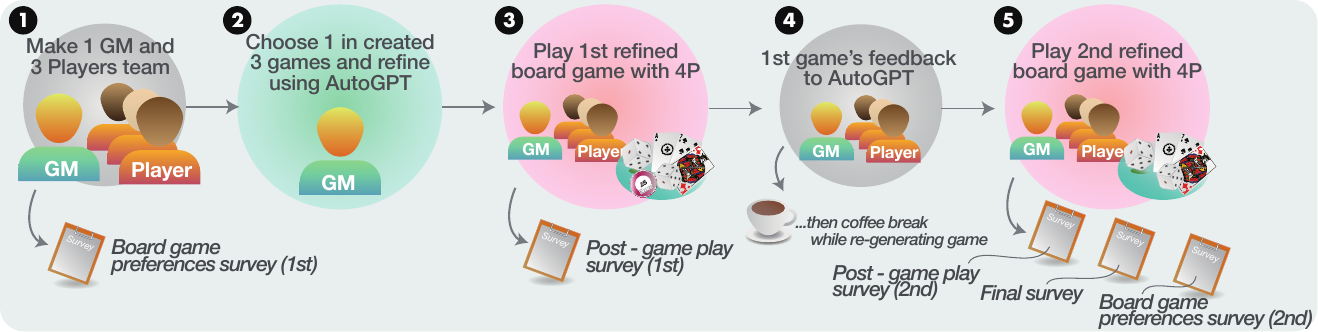}
  \caption{The user study design: Step1-board game preference survey Step2-select a board game generated and give first feedback to the AutoGPT system. Step3-playtest the game the first time and answer post-game play survey. Step4-game master gives second feedback to the system. Step5-playtest the game the second time and answer post-game play survey, final survey and board game preference survey.}
  \Description{The user study design: Step1-board game preference survey Step2-select a board game generated and give first feedback to the AutoGPT system. Step3-playtest the game the first time and answer post-game play survey. Step4-game master gives second feedback to the system. Step5-playtest the game the second time and answer post-game play survey, final survey and board game preference survey.}
  \label{us}
\end{figure*}

\subsection{User Study Design and Methodology}

The user study was designed as a two-game experiment. Each game consisted of a playtest and a survey. Participants wrote feedback in the format of listing "problems" and "requirements" as a set in their native language, and the study conductor then translated it into English for the AutoGPT system. The overall view of the user study is explained in Fig.~\ref{us}.

At the beginning and end of each phase, participants completed a survey designed to gather quantitative and qualitative data on the AutoGPT system's usability, the playability of the generated games, and the overall enjoyment and engagement of the games.

In the user study, 12 participants were grouped into three teams of four: one volunteer Game Master (GM) and three players per team. The participants, with ages ranging from 20 to 27 years and a balanced representation of both genders, were recruited from the University of Tsukuba. Participants had varied board game experiences, ranging from less than a year to over a decade, with play frequencies from a few times a year to once a month. The GMs, responsible for guiding sessions, chose a board game from three generated rules, provided system feedback, and supervised playtesting. The players were responsible for playtesting the game, providing feedback, and participating in the discussion.

\subsection{Data Collection}

Data was collected through board game preference surveys, post-gameplay surveys, final surveys, Creativity Support Index (CSI)~\cite{10.1145/1640233.1640255} for Game Masters (GMs), and user study conductor observations. This approach allowed us to gather insights into the usability, playability, and enjoyment of the AutoGPT-generated games.
Surveys were administered to gauge participants' initial expectations and subsequent impressions. Post-gameplay surveys covered various aspects of game experience, while the final survey allowed participants to reflect on their overall experience. The CSI~\cite{10.1145/1640233.1640255} was used to evaluate the level of creativity support provided by the AutoGPT system to the GMs.
GM's feedback, survey responses, and conductor observations contributed to the data collection regarding the system's usability and game development process.

%% file: camera-ready_sections/5_result-discuss.tex
\begin{figure*}[ht]
  \centering
  \includegraphics[width=\linewidth]{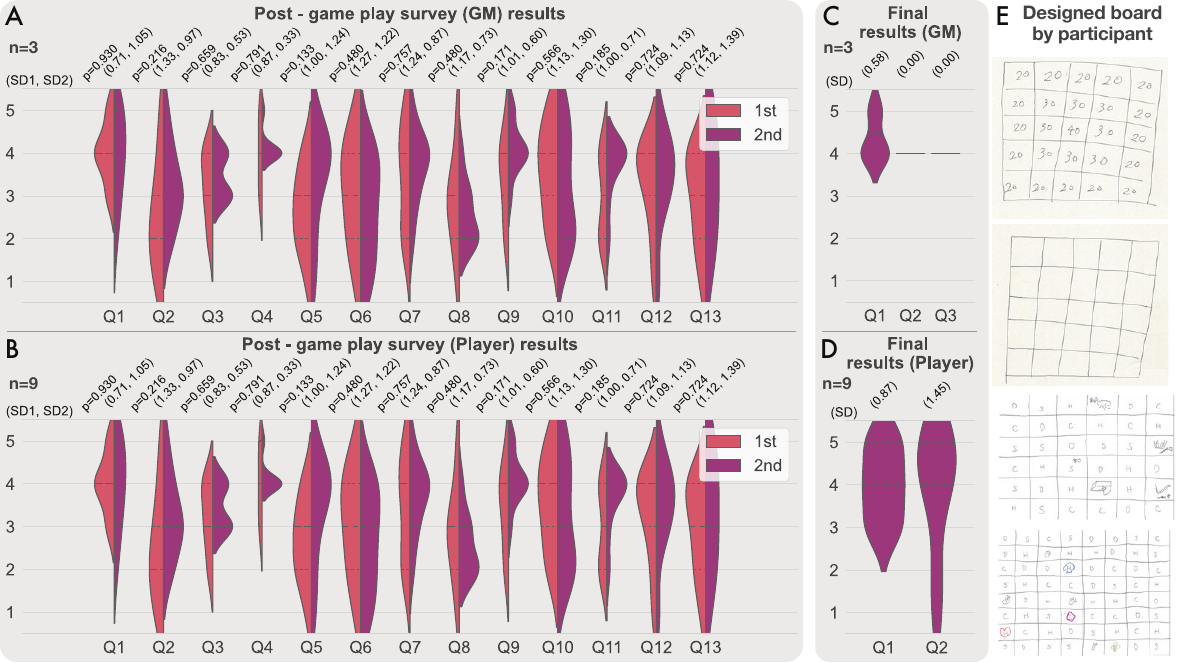}
  \caption{(A) Quantitative results of “post-gameplay survey” GM- Q1 Overall, the game was what I expected. Q2 the game rule was understandable. Q3 the rule documents were written for easy to graphic design. Q4 The game had enough strategic elements. Q5 the length of the game was appropriate. Q6 the rules of the game were fair. Q7 the game had the originality. Q8 the choice of theme for the game was what I expected. Q9 there was appropriate interaction with other players. Q10 I could prepare a contingent event. Q11 There was adequate player control elements in the game. (B) Player- Q1 I enjoyed. Q2 the game rule was understandable. Q3 the rule documents were written for easy to graphic design. Q4 the length of the game was appropriate. Q5 The game had enough strategic elements. Q6 the rules of the game were fair. Q7 the game had the originality. Q8 the theme of the game was great. Q9 The game had interaction with other players. Q10 The game had contingent events. Q11 the game had adequate control over what I could control in the game. Q12 I want to play the game again. Q13 I want to play games with friends and family. (C) Quantitative results of “final survey” GM- Q1 The board games output by GPT were fascinating. Q2 the board games output by GPT were as expected. Q3 comparing the first and second sessions, GPT's improved board game was as expected. (D) Player- Q1 the board games were fascinating Q2 Comparing the first and second sessions, the board game was improved. (E)Board of the board game drawn by the participant for playtest.
}
\label{b}
\end{figure*}

\begin{figure*}[ht]
  \centering
  \includegraphics[width=\linewidth]{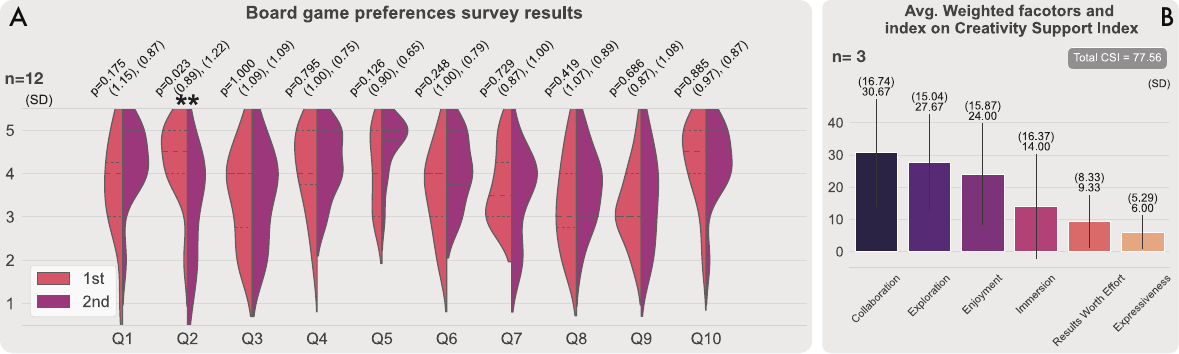}
  \caption{(A)quantitative results of “board game preference survey” Questions:Q1 Concepts Q2 Board graphic design Q3 Game themes Q4 Game strategy Q5 Communications Q6 speed of progression Q7 Easy rule Q8 Randomness Q9 Controllable many objects Q10 Fairness rule. p ** < 0.05. and (B) Average of weighted factors and total CSI result.}
  \Description{(A)“board game preference survey” Questions:Q1 Concepts Q2 Board graphic design Q3 Game themes Q4 Game strategy Q5 Communications Q6 speed of progression Q7 Easy rule Q8 Randomness Q9 Controllable many objects Q10 Fairness rule. p ** < 0.05. and (B) Avg. CSI test result.}
  \label{a}
\end{figure*}

Overall Assessment of the AutoGPT System: The results of the final survey showed that participants provided positive evaluations of the AutoGPT board game creation system. In the Game Master (GM) category, high scores were recorded for Q1 (Median: 4.00, SD: 0.58), Q2 (Median: 4.00, SD: 0.00), and Q3 (Median: 4.00, SD: 0.00). Similarly, in the player category, high scores were observed for Q1 (Median: 4.00, SD: 0.87) and Q2 (Median: 4.00, SD: 1.45) in the final survey. These high scores indicate that the generated games met participants' expectations and that the AutoGPT system was effective in creating board games that were generally well-received (Fig.~\ref{b}).

The system scored an average of 77.58 (n=3) on CSI scores ranging from 0 (worst) to 100 (best). The highest factor was Collaboration with an average of 30.67, followed by Exploration with an average of 27.67. Collaboration evaluates if the system allowed other people to work with the user easily, and Exploration evaluates if the user could explore many different ideas, options, designs, or outcomes. Expressiveness had the lowest score of 5.29, evaluating if the user was able to be very expressive and creative while doing the activity. This result is supported by a comment from the final GM survey, where the respondent was asked to identify which aspects of the system they found convenient or supportive and any difficulties they encountered: "I can't predict what kind of instructions make what kind of difference."

Comprehensibility of Game Rules: In the post-gameplay survey, both GMs and players reported increased median scores for the understandability of game rules in the second game compared to the first game. GMs showed an increase in median score from 2.00 in the first game to 3.00 in the second game (SD1: 1.00, SD2: 1.00), and players displayed an increase in the median score from 2.00 to 3.00 as well (SD1: 1.33, SD2: 0.97). However, this improvement could be attributed to participants playing the game twice with similar rules. It remains unclear whether the generated game itself enhanced its comprehensibility (Fig.~\ref{b}). 
Several participants provided qualitative comments, with one noting, "It would be better if the content was a bit more detailed." The lack of clarity in the generated rules was also identified as a potential factor affecting comprehensibility.

Strategic Elements and Fairness: The evaluation of strategic elements revealed a discrepancy between GM and player perceptions, with GMs reporting a higher score in the second game, increasing the median score from 3.00 to 4.00 (SD1: 1.53, SD2: 0.58). In contrast, players' evaluations remained unchanged, with their median score staying at 3.00 in both the first and second game (SD1: 1.00, SD2: 1.27). This suggests that the improvement in strategic elements, based on GM input and expectations, did not translate into an enhanced experience for the players. Similarly, evaluations of fairness improved for GMs, with an increase in the median score from 2.00 in the first game to 4.00 in the second game (SD1: 1.73, SD2: 2.08); however, player evaluations did not exhibit a significant change between the first and second games (Median1: 3.00, Median2: 3.00, SD1: 1.27, SD2: 1.23). This suggests that only GMs were satisfied with the improvements in fairness while players did not share the same sentiment (Fig.~\ref{b}). 
Player feedback, such as "The game balance isn't fully adjusted," supports this observation. Even with GMs having the authority to adjust rules, ensuring universally accepted fairness was challenging.

Originality and Game Theme: The AutoGPT system successfully delivered novel and original game concepts, as evidenced by consistently high scores provided by both GMs and players. In the post-gameplay survey, GMs recorded median scores of 4.00 for both games (SD1: 0.58, SD2: 0.00), and players reported median scores of 4.00 for both games as well (SD1: 1.24, SD2: 0.87). In contrast, evaluations of game theme showed differing opinions between GMs, who perceived a positive improvement, and players whose assessment declined in the second game (Fig.~\ref{b}). GMs offered qualitative comments such as, "It's unpredictable how instructions will alter the game," and "It brings ideas humans might not immediately consider." These remarks highlight the system's unique ability to introduce originality in game creation.

Contingency and Board Graphic Design: The decrease in player evaluations of contingency in the post-gameplay survey, with the median score dropping from 4.00 in the first game to 3.00 in the second game (SD1: 1.13, SD2: 1.30), and the increased importance placed on contingency in the board game preference survey indicate that contingent or random events play a crucial role in the overall enjoyment of generated board games (Fig.~\ref{a}).

Furthermore, high-quality graphic design may not be as essential to board game enjoyment as initially thought (Fig.~\ref{b}A,B,D). The evaluation of board graphic design in the "board game preference survey" (Q2) changed, with the median score decreasing from 4.50 (SD1: 0.89) to 4.00 (SD2: 1.22)(Fig.~\ref{a}). Despite its simplicity, the hand-drawn board provided an engaging gaming experience. This finding aligns with Kieran Hicks et al.'s observation that visual embellishments do not always improve the user experience~\cite{10.1145/3311350.3347171}.

%% file: camera-ready_sections/6_limitation-future.tex
Our study of the AutoGPT system for board game creation showed promising results. However, there are some limitations to address, and we suggest potential areas for future research.
First, the system had difficulty creating complete game board designs, often resulting in issues with overlapping components and missing elements. Future work could focus on improving the graphical capabilities of the AutoGPT system for better board designs.
Second, sometimes the system generated games with contradictory rules or materials not allowed by the given constraints. To make the games more playable, future research should focus on refining the system's ability to understand constraints and provide clear rules.
Lastly, the limited working memory of the AutoGPT system prevented it from researching existing board games thoroughly. Enhancing its working memory by developing database system could help the system better understand existing games and create more engaging designs.

%% file: camera-ready_sections/7_conclusion.tex
In conclusion, our study demonstrated the potential of the AutoGPT system in generating engaging and original board games. By addressing limitations and refining the system's capabilities, it can better match players' expectations and deliver satisfying gaming experiences. Future work can focus on enhancing the system for various game types, contributing to the advancement of human-computer interaction in the gaming industry. Overall, the AutoGPT system shows promise in revolutionizing the game creation process and offering enjoyable, well-rounded board games for diverse audiences and gaming preferences.

%% file: main.bbl

\begin{thebibliography}{18}


\ifx \showCODEN    \undefined \def \showCODEN     #1{\unskip}     \fi
\ifx \showDOI      \undefined \def \showDOI       #1{#1}\fi
\ifx \showISBNx    \undefined \def \showISBNx     #1{\unskip}     \fi
\ifx \showISBNxiii \undefined \def \showISBNxiii  #1{\unskip}     \fi
\ifx \showISSN     \undefined \def \showISSN      #1{\unskip}     \fi
\ifx \showLCCN     \undefined \def \showLCCN      #1{\unskip}     \fi
\ifx \shownote     \undefined \def \shownote      #1{#1}          \fi
\ifx \showarticletitle \undefined \def \showarticletitle #1{#1}   \fi
\ifx \showURL      \undefined \def \showURL       {\relax}        \fi
\providecommand\bibfield[2]{#2}
\providecommand\bibinfo[2]{#2}
\providecommand\natexlab[1]{#1}
\providecommand\showeprint[2][]{arXiv:#2}

\bibitem[Araújo et~al\mbox{.}(2019)]%
        {inbook}
\bibfield{author}{\bibinfo{person}{Carlos Araújo}, \bibinfo{person}{Ivon Santos}, \bibinfo{person}{E.D. Canedo}, {and} \bibinfo{person}{Aletéia Araújo}.} \bibinfo{year}{2019}\natexlab{}.
\newblock \bibinfo{booktitle}{\emph{Design Thinking Versus Design Sprint: A Comparative Study}}.
\newblock 291--306 pages.
\newblock
\showISBNx{978-3-030-23569-7}
\urldef\tempurl%
\url{https://doi.org/10.1007/978-3-030-23570-3_22}
\showDOI{\tempurl}


\bibitem[Carroll et~al\mbox{.}(2009)]%
        {10.1145/1640233.1640255}
\bibfield{author}{\bibinfo{person}{Erin~A. Carroll}, \bibinfo{person}{Celine Latulipe}, \bibinfo{person}{Richard Fung}, {and} \bibinfo{person}{Michael Terry}.} \bibinfo{year}{2009}\natexlab{}.
\newblock \showarticletitle{Creativity Factor Evaluation: Towards a Standardized Survey Metric for Creativity Support}. In \bibinfo{booktitle}{\emph{Proceedings of the Seventh ACM Conference on Creativity and Cognition}} (Berkeley, California, USA) \emph{(\bibinfo{series}{C\&C '09})}. \bibinfo{publisher}{Association for Computing Machinery}, \bibinfo{address}{New York, NY, USA}, \bibinfo{pages}{127–136}.
\newblock
\showISBNx{9781605588650}
\urldef\tempurl%
\url{https://doi.org/10.1145/1640233.1640255}
\showDOI{\tempurl}


\bibitem[Chaslot et~al\mbox{.}(2008)]%
        {chaslot2008monte}
\bibfield{author}{\bibinfo{person}{Guillaume Chaslot}, \bibinfo{person}{Sander Bakkes}, \bibinfo{person}{Istvan Szita}, {and} \bibinfo{person}{Pieter Spronck}.} \bibinfo{year}{2008}\natexlab{}.
\newblock \showarticletitle{Monte-carlo tree search: A new framework for game ai}. In \bibinfo{booktitle}{\emph{Proceedings of the AAAI Conference on Artificial Intelligence and Interactive Digital Entertainment}}, Vol.~\bibinfo{volume}{4}. \bibinfo{pages}{216--217}.
\newblock


\bibitem[{De Giacomo} et~al\mbox{.}(2020)]%
        {ludii}
\bibfield{editor}{\bibinfo{person}{Giuseppe {De Giacomo}}, \bibinfo{person}{Alejandro Catala}, \bibinfo{person}{Bistra Dilkina}, \bibinfo{person}{Michela Milano}, \bibinfo{person}{Sen{\'e}n Barro}, \bibinfo{person}{Alberto Bugar{\'i}n}, {and} \bibinfo{person}{J{\'e}r{\^o}me Lang}} (Eds.). \bibinfo{year}{2020}\natexlab{}.
\newblock \showarticletitle{Ludii - The Ludemic General Game System}.
\newblock   \bibinfo{volume}{325} (\bibinfo{year}{2020}), \bibinfo{pages}{411--418}.
\newblock
\showISBNx{978-1-64368-100-9}
\urldef\tempurl%
\url{https://doi.org/10.3233/FAIA200120}
\showDOI{\tempurl}


\bibitem[Eck et~al\mbox{.}(2017)]%
        {eck2017leveling}
\bibfield{author}{\bibinfo{person}{Van Eck}, \bibinfo{person}{R. N.}, \bibinfo{person}{V.~J. Shute}, {and} \bibinfo{person}{L.~P. Rieber}.} \bibinfo{year}{2017}\natexlab{}.
\newblock \showarticletitle{Leveling up: {Game} design research and practice for instructional designers}.
\newblock \bibinfo{publisher}{Pearson}, \bibinfo{address}{New York, NY}, \bibinfo{pages}{227--285}.
\newblock


\bibitem[Hicks et~al\mbox{.}(2019)]%
        {10.1145/3311350.3347171}
\bibfield{author}{\bibinfo{person}{Kieran Hicks}, \bibinfo{person}{Kathrin Gerling}, \bibinfo{person}{Patrick Dickinson}, {and} \bibinfo{person}{Vero Vanden~Abeele}.} \bibinfo{year}{2019}\natexlab{}.
\newblock \showarticletitle{Juicy Game Design: Understanding the Impact of Visual Embellishments on Player Experience}. In \bibinfo{booktitle}{\emph{Proceedings of the Annual Symposium on Computer-Human Interaction in Play}} (Barcelona, Spain) \emph{(\bibinfo{series}{CHI PLAY '19})}. \bibinfo{publisher}{Association for Computing Machinery}, \bibinfo{address}{New York, NY, USA}, \bibinfo{pages}{185–197}.
\newblock
\showISBNx{9781450366885}
\urldef\tempurl%
\url{https://doi.org/10.1145/3311350.3347171}
\showDOI{\tempurl}


\bibitem[Khalifa et~al\mbox{.}(2019)]%
        {khalifa2019general}
\bibfield{author}{\bibinfo{person}{Ahmed Khalifa}, \bibinfo{person}{Michael~Cerny Green}, \bibinfo{person}{Diego Perez-Liebana}, {and} \bibinfo{person}{Julian Togelius}.} \bibinfo{year}{2019}\natexlab{}.
\newblock \bibinfo{title}{General Video Game Rule Generation}.
\newblock
\newblock
\showeprint[arxiv]{1906.05160}~[cs.AI]


\bibitem[Knapp et~al\mbox{.}(2016)]%
        {knapp2016sprint}
\bibfield{author}{\bibinfo{person}{J. Knapp}, \bibinfo{person}{J. Zeratsky}, {and} \bibinfo{person}{B. Kowitz}.} \bibinfo{year}{2016}\natexlab{}.
\newblock \bibinfo{booktitle}{\emph{Sprint: How to Solve Big Problems and Test New Ideas in Just Five Days}}.
\newblock \bibinfo{publisher}{Simon \& Schuster}.
\newblock
\showISBNx{9781501121777}
\urldef\tempurl%
\url{https://books.google.co.jp/books?id=rV0JCgAAQBAJ}
\showURL{%
\tempurl}


\bibitem[Kowalski et~al\mbox{.}(2019)]%
        {kowalski2019regular}
\bibfield{author}{\bibinfo{person}{Jakub Kowalski}, \bibinfo{person}{Maksymilian Mika}, \bibinfo{person}{Jakub Sutowicz}, {and} \bibinfo{person}{Marek Szyku{\l}a}.} \bibinfo{year}{2019}\natexlab{}.
\newblock \showarticletitle{Regular boardgames}. In \bibinfo{booktitle}{\emph{Proceedings of the AAAI Conference on Artificial Intelligence}}, Vol.~\bibinfo{volume}{33}. \bibinfo{pages}{1699--1706}.
\newblock


\bibitem[Loper and Bird(2002)]%
        {loper2002nltk}
\bibfield{author}{\bibinfo{person}{Edward Loper} {and} \bibinfo{person}{Steven Bird}.} \bibinfo{year}{2002}\natexlab{}.
\newblock \showarticletitle{Nltk: The natural language toolkit}.
\newblock \bibinfo{journal}{\emph{arXiv preprint cs/0205028}} (\bibinfo{year}{2002}).
\newblock


\bibitem[Mochocki and Milewski(2018)]%
        {book}
\bibfield{author}{\bibinfo{person}{Michal Mochocki} {and} \bibinfo{person}{Piotr Milewski}.} \bibinfo{year}{2018}\natexlab{}.
\newblock \bibinfo{booktitle}{\emph{Game Design Curriculum White Paper 2.0}}.
\newblock
\showISBNx{978-83-952601-0-0}


\bibitem[Osone et~al\mbox{.}(2021)]%
        {buncho}
\bibfield{author}{\bibinfo{person}{Hiroyuki Osone}, \bibinfo{person}{Jun-Li Lu}, {and} \bibinfo{person}{Yoichi Ochiai}.} \bibinfo{year}{2021}\natexlab{}.
\newblock \showarticletitle{BunCho: AI Supported Story Co-Creation via Unsupervised Multitask Learning to Increase Writers’ Creativity in Japanese}. In \bibinfo{booktitle}{\emph{Extended Abstracts of the 2021 CHI Conference on Human Factors in Computing Systems}} (Yokohama, Japan) \emph{(\bibinfo{series}{CHI EA '21})}. \bibinfo{publisher}{Association for Computing Machinery}, \bibinfo{address}{New York, NY, USA}, Article \bibinfo{articleno}{19}, \bibinfo{numpages}{10}~pages.
\newblock
\showISBNx{9781450380959}
\urldef\tempurl%
\url{https://doi.org/10.1145/3411763.3450391}
\showDOI{\tempurl}


\bibitem[Significant-Gravitas(2023)]%
        {Auto-GPT}
\bibfield{author}{\bibinfo{person}{Significant-Gravitas}.} \bibinfo{year}{2023}\natexlab{}.
\newblock \bibinfo{title}{Significant-Gravitas/Auto-GPT}.
\newblock
\newblock
\urldef\tempurl%
\url{https://github.com/Significant-Gravitas/Auto-GPT}
\showURL{%
\tempurl}


\bibitem[Singhal et~al\mbox{.}(2001)]%
        {singhal2001modern}
\bibfield{author}{\bibinfo{person}{Amit Singhal} {et~al\mbox{.}}} \bibinfo{year}{2001}\natexlab{}.
\newblock \showarticletitle{Modern information retrieval: A brief overview}.
\newblock \bibinfo{journal}{\emph{IEEE Data Eng. Bull.}} \bibinfo{volume}{24}, \bibinfo{number}{4} (\bibinfo{year}{2001}), \bibinfo{pages}{35--43}.
\newblock


\bibitem[Thielscher(2017)]%
        {thielscher2017gdl}
\bibfield{author}{\bibinfo{person}{Michael Thielscher}.} \bibinfo{year}{2017}\natexlab{}.
\newblock \showarticletitle{GDL-III: A description language for epistemic general game playing}. In \bibinfo{booktitle}{\emph{The IJCAI-16 workshop on general game playing}}. \bibinfo{pages}{31}.
\newblock


\bibitem[Todd et~al\mbox{.}(2023)]%
        {10.1145/3582437.3587211}
\bibfield{author}{\bibinfo{person}{Graham Todd}, \bibinfo{person}{Sam Earle}, \bibinfo{person}{Muhammad~Umair Nasir}, \bibinfo{person}{Michael~Cerny Green}, {and} \bibinfo{person}{Julian Togelius}.} \bibinfo{year}{2023}\natexlab{}.
\newblock \showarticletitle{Level Generation Through Large Language Models}. In \bibinfo{booktitle}{\emph{Proceedings of the 18th International Conference on the Foundations of Digital Games}} (Lisbon, Portugal) \emph{(\bibinfo{series}{FDG '23})}. \bibinfo{publisher}{Association for Computing Machinery}, \bibinfo{address}{New York, NY, USA}, Article \bibinfo{articleno}{70}, \bibinfo{numpages}{8}~pages.
\newblock
\showISBNx{9781450398558}
\urldef\tempurl%
\url{https://doi.org/10.1145/3582437.3587211}
\showDOI{\tempurl}


\bibitem[Yang et~al\mbox{.}(2023b)]%
        {yang2023autogpt}
\bibfield{author}{\bibinfo{person}{Hui Yang}, \bibinfo{person}{Sifu Yue}, {and} \bibinfo{person}{Yunzhong He}.} \bibinfo{year}{2023}\natexlab{b}.
\newblock \bibinfo{title}{Auto-GPT for Online Decision Making: Benchmarks and Additional Opinions}.
\newblock
\newblock
\showeprint[arxiv]{2306.02224}~[cs.AI]


\bibitem[Yang et~al\mbox{.}(2023a)]%
        {yang2023harnessing}
\bibfield{author}{\bibinfo{person}{Jingfeng Yang}, \bibinfo{person}{Hongye Jin}, \bibinfo{person}{Ruixiang Tang}, \bibinfo{person}{Xiaotian Han}, \bibinfo{person}{Qizhang Feng}, \bibinfo{person}{Haoming Jiang}, \bibinfo{person}{Bing Yin}, {and} \bibinfo{person}{Xia Hu}.} \bibinfo{year}{2023}\natexlab{a}.
\newblock \bibinfo{title}{Harnessing the Power of LLMs in Practice: A Survey on ChatGPT and Beyond}.
\newblock
\newblock
\showeprint[arxiv]{2304.13712}~[cs.CL]


\end{thebibliography}
